\documentclass[onecolumn,showpacs,aps,amsfonts,floatfix]{revtex4-1}
\usepackage{graphicx}
\usepackage{times}
\usepackage{color}
\usepackage[T1]{fontenc}
\usepackage{amssymb}
\usepackage{amsmath}

\usepackage[latin1]{inputenc}

\DeclareMathOperator{\Tr}{Tr}
\newcommand{\be}{\begin{equation}}
\newcommand{\ee}{\end{equation}}
\newcommand{\br}{\begin{eqnarray}}
\newcommand{\er}{\end{eqnarray}}

\newcommand{\bd}{\begin{displaymath}}
\newcommand{\ed}{\end{displaymath}}

\newcommand{\bfig}{\begin{figure}}
\newcommand{\efig}{\end{figure}}

\def\3cdot{\cdot \cdot \cdot}

\def\om0{\omega _0}
\def\Om0{\Omega _0}

\def\text#1{{\rm{#1}}}

\def\->{\rightarrow}
\def\=>{\Rightarrow}
\def\-->{\longrightarrow}
\def\==>{\Longrightarrow}

\def\dag{\dagger}

\def\pr{^\prime}
\def\pr2{^{\prime\prime}}

\def\bfig{\begin{figure}}
\def\efig{\end{figure}}

\begin{document}

\title{Temperature Measurement and Phonon Number Statistics of a Nanoelectromechanical Resonator}

\author{O. P. de S\'a Neto$^{1,2}$\email{olimpioqedc@gmail.com}, M. C. de Oliveira$^{2}$ and G. J. Milburn$^3$}

\affiliation{$^1$ Coordenação de Ciência da Computação,
Universidade Estadual do Piau\'i, CEP: 64202220, Parnaíba, Piau\'i, Brazil.}

\affiliation{$^2$ Instituto de F\'\i sica ``Gleb Wataghin'',
 Universidade Estadual de Campinas, 13083-970, Campinas, S\~ao Paulo, Brazil.}

\affiliation{$^3$ Centre for Engineered Quantum Systems,  School of Mathematics and Physics., University of Queensland, QLD 4072, Brisbane, Australia.}
\date{\today}

\begin{abstract}
Measuring thermodynamic quantities can be easy or not, depending on the system that is being studied. For a macroscopic object, measuring temperatures can be as simple as measuring how much a column of mercury rises when in contact with the object. At the small scale of quantum electromechanical systems, such simple methods are not available and invariably detection processes disturb the system state. Here we propose a method for measuring the temperature on a suspended semiconductor membrane clamped at both ends. In this method, the membrane is mediating a capacitive coupling between two transmission line resonators (TLR). The first TLR has a strong dispersion, that is, its decaying rate is larger than its drive, and its role is to pump in a pulsed way the interaction between the membrane and the second TLR. By averaging the pulsed measurements of the quadrature of the second TLR we show how the temperature of the membrane can be determined. Moreover the statistical description of the state of the membrane, which is directly accessed in this approach is significantly improved by the addition of a Josephson Junction coupled to the second TLR.
\end{abstract}

\maketitle

\section{Introduction}
Electromechanical systems are devices which couple mechanical displacement and electrostatic interactions. Measuring physical properties of such a device at macroscopic scales is relatively easy - Coulomb in his famous torsion balance attached charged metal spheres to rods and threads and visually measured the torsion produced by the electrostatic interaction between the spheres. The measurement of the torsion allowed him to determine the force acting on the spheres \cite{World,Ptoday}. At the nanoscale however \cite{bohr1949discussion}, the movement of a nanoelectromechanical system (NEMS) cannot be observed directly. A further complication is that at those scales a quantum description of the system is invariably necessary.

The specific NEMS we are interested in is a suspended semiconductor membrane clamped at both ends, which will be oscillating due to coupling to the thermal modes of the clamps. The smaller is the mechanical element of the NEMS, the stronger is the coupling to the thermal modes of the reservoirs that clamp them at the extremities, and usually some enginnered structures must be employed in order to decrease its effects \cite{Chan2011}. This oscillator can be coupled electrostatically to other devices, allowing the transduction of the movement as electric signals {(See Ref. \cite{asp2014} for a general review)}. Previously schemes to measure the quadrature phase amplitude \cite{Tlrnems} and to observe the quantum of thermal conductance \cite{Preport} have been proposed. 
It is particularly relevant that the detection of the NEMS movement can give direct access to its temperature, a fundamental physical quantity \cite{Chang2007,Stace2010}. For the NEMS temperature measurement, usually the area under the noise power spectrum of the displacement amplitude transduced signal is used as it gives directly the mean phonon number in the steady state (See \cite{milburn2011introduction,Chan2011} for example).  
However, one could ask on how to access the temperature of the mechanical resonator by means of a non-demolition detection scheme. Indeed, some previous discussion on non-demolition detection in the accessment of temperature has been given \cite{PhysRevLett.94.030402,PhysRevLett.95.248901}. {Also a discussion on back-action-evading detection and squeezing generation in mechanical resonators through coupling to radiation field were given in \cite{clerk2008} as well as the outstanding implementation for movement detection given in \cite{hetzberg2009}.}

In this work we describe a circuit in which we can perform {repeated} quantum nondemolition measurement sof the expected value of the number of phonons of the NEMS, and show an analysis of the statistics of the number of phonons in this circuit. Through this analysis we show how the temperature of the semiconductor membrane can be directly accessed. Finally, we show that by discretizing the measurement of the expected value we can increase its accuracy in a pulsed linear radiation detection process \cite{6,Bozyigit2010}. 
\begin{figure}[h]
\includegraphics[scale=0.3]{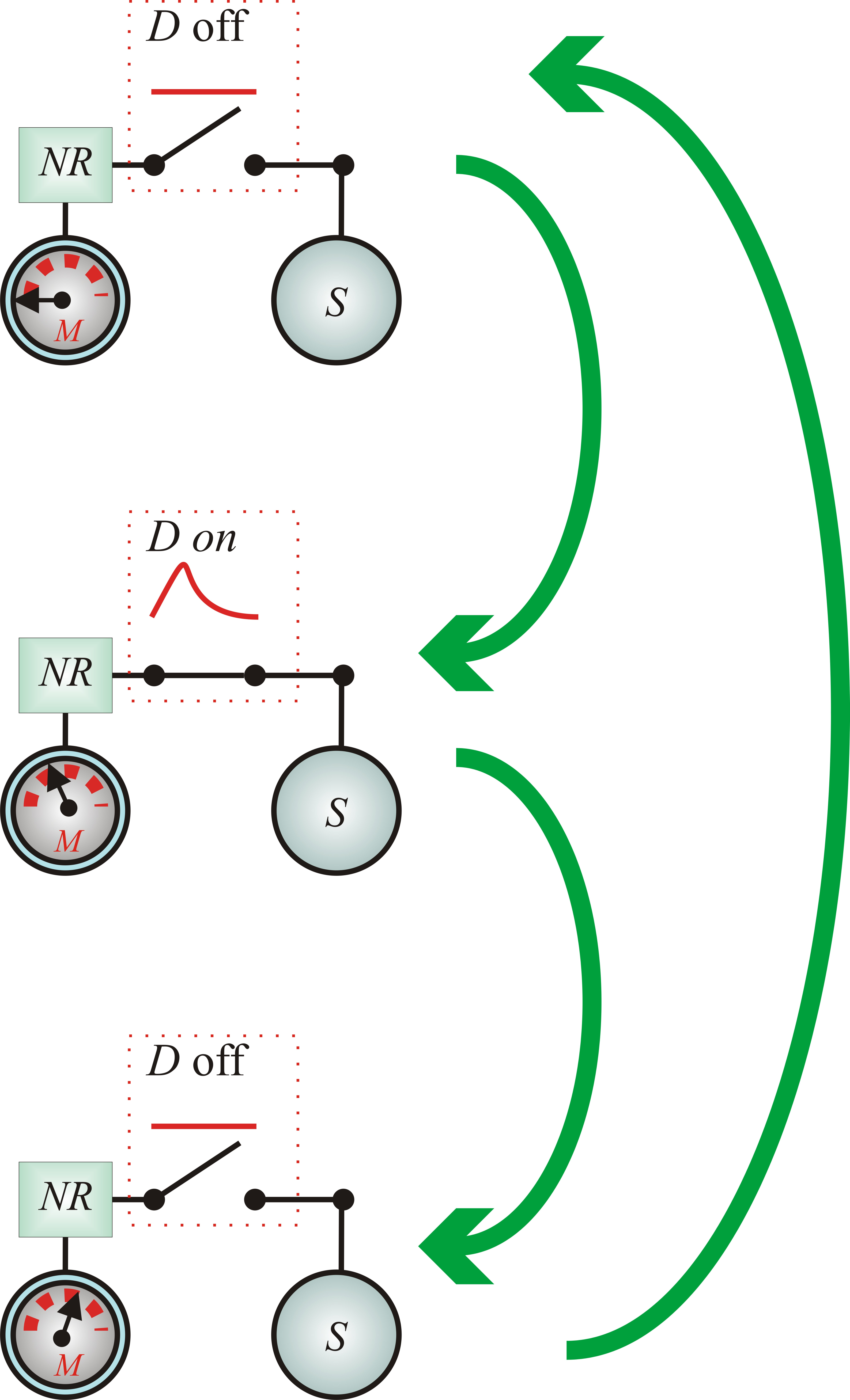}
\caption{Diagram for a pulsed quantum nondemolition measurement. The drive ``\textit{D}'' pulses the interaction between the system ``\textit{S}'' and the measuring device ``\textit{M}''. The noise reducer ``\textit{NR}'' lowers the fluctuations of the meter in each measurement. This cycle, is repeated many times, always following the order: state preparation $\rightarrow$ coherent control $\rightarrow$ quantum measurement.
}
\label{diagram}
\end{figure}

In order to perform a measurement on a small oscillator, in a quantum regime \cite{World,Ptoday,Lahaye,hetzberg2009}, a quantum nondemolition measurement is necessary. In this type of measurement, the coupling with the meter is such that it does not disturbs the non-demolition observable of interest. Also, because these measurements are probabilistic by nature, it is also important to improve the precision of the meter so that at the end of many measurements there is a significant accuracy in the calculated average. This cyclic sequence of non continuous measurements in time is shown on the diagram of Fig. \ref{diagram}.

\section{Results}

\begin{figure*}[t]
\includegraphics[scale=0.55]{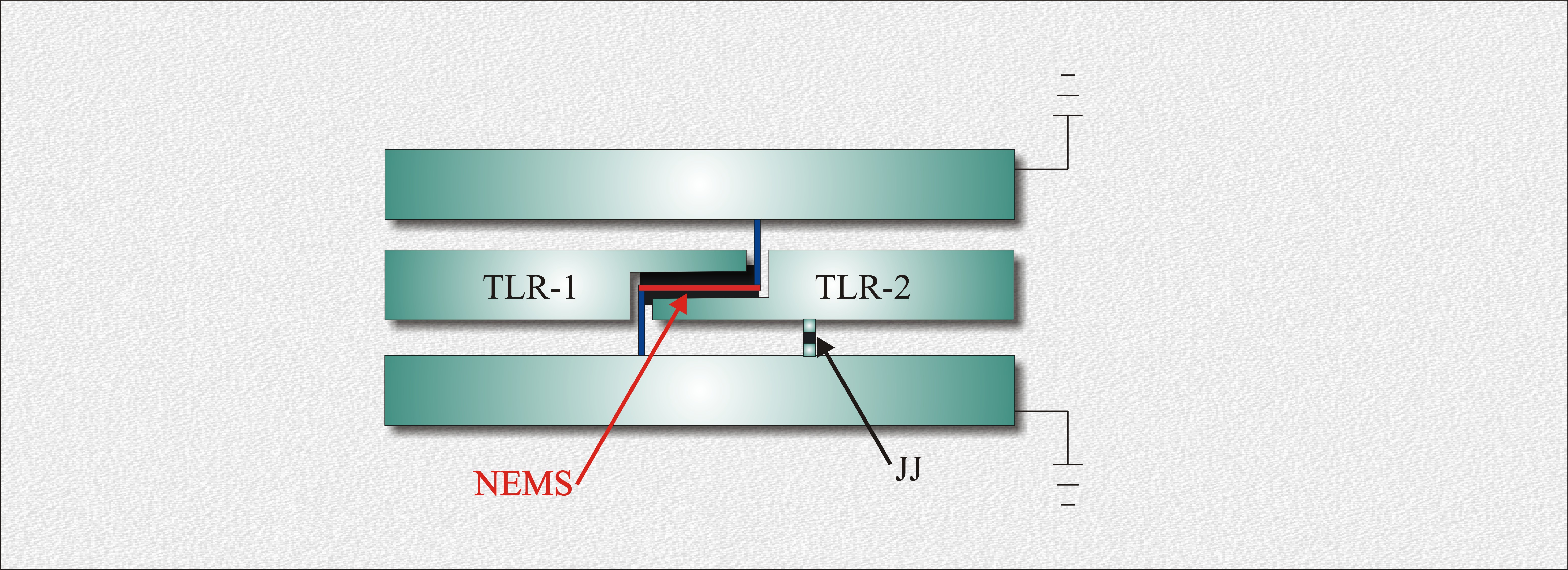}\\(a)\\
\includegraphics[scale=0.55]{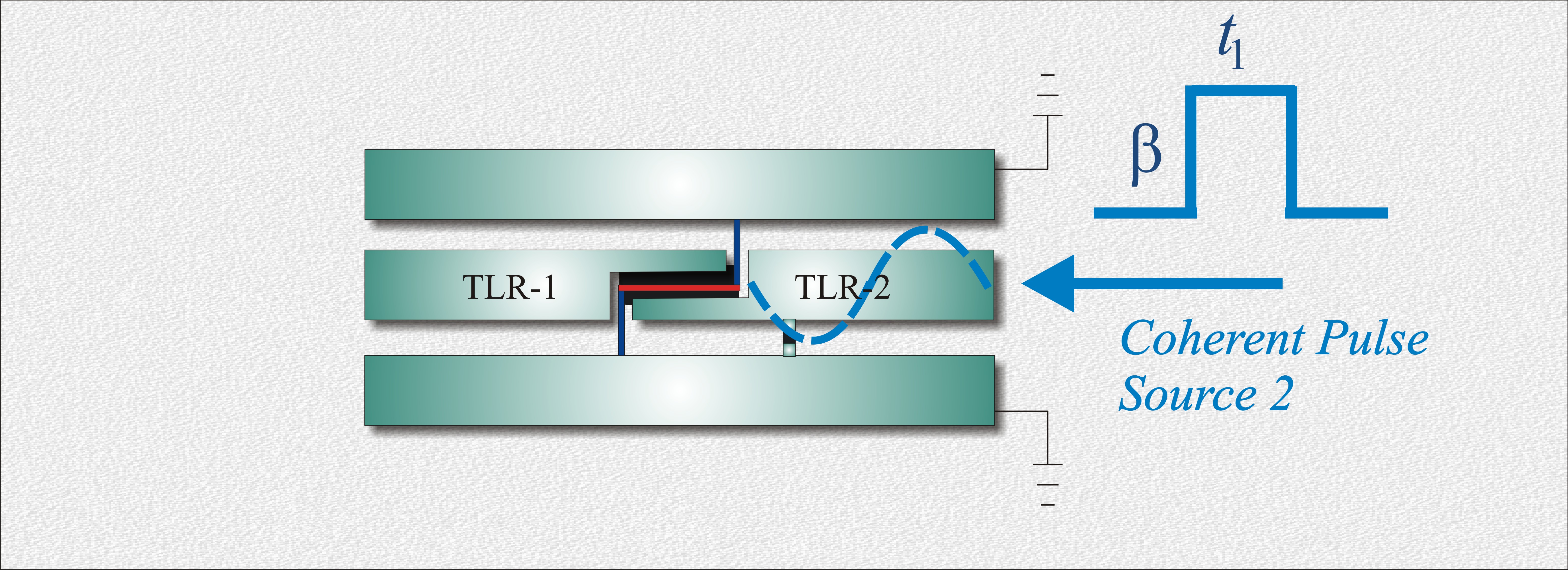}\\(b)\\
\includegraphics[scale=0.55]{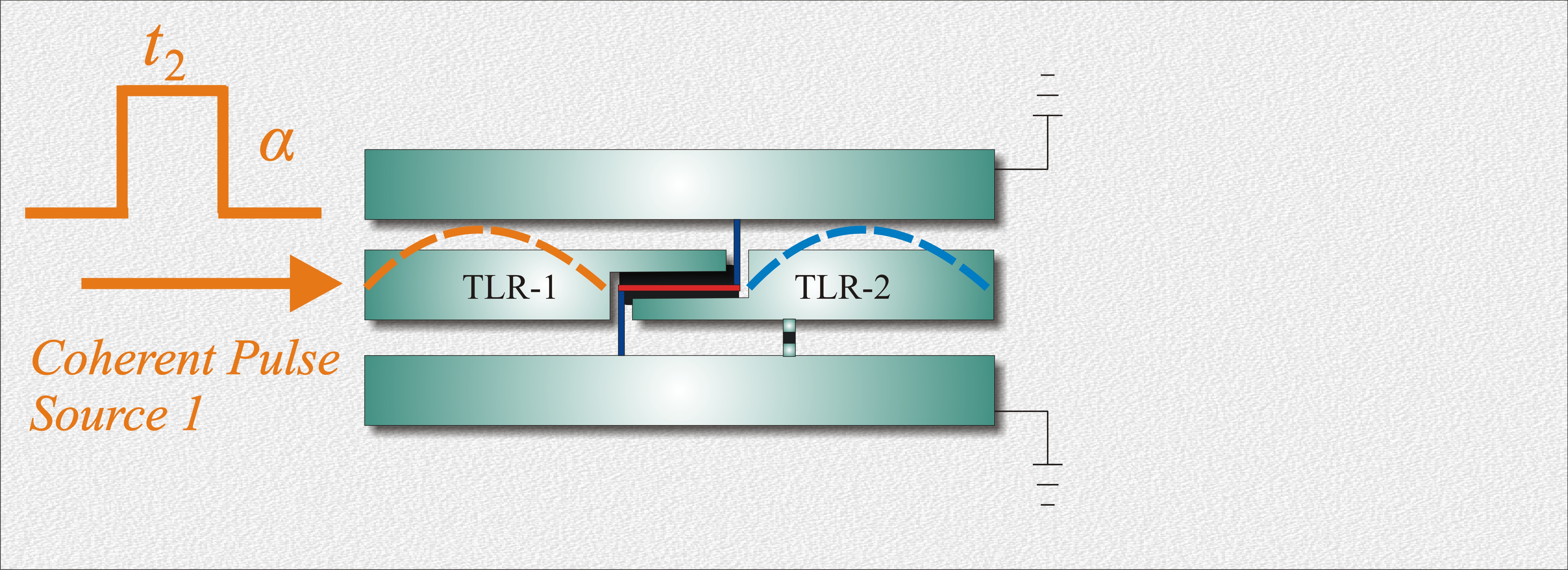}\\(c)\\
\includegraphics[scale=0.55]{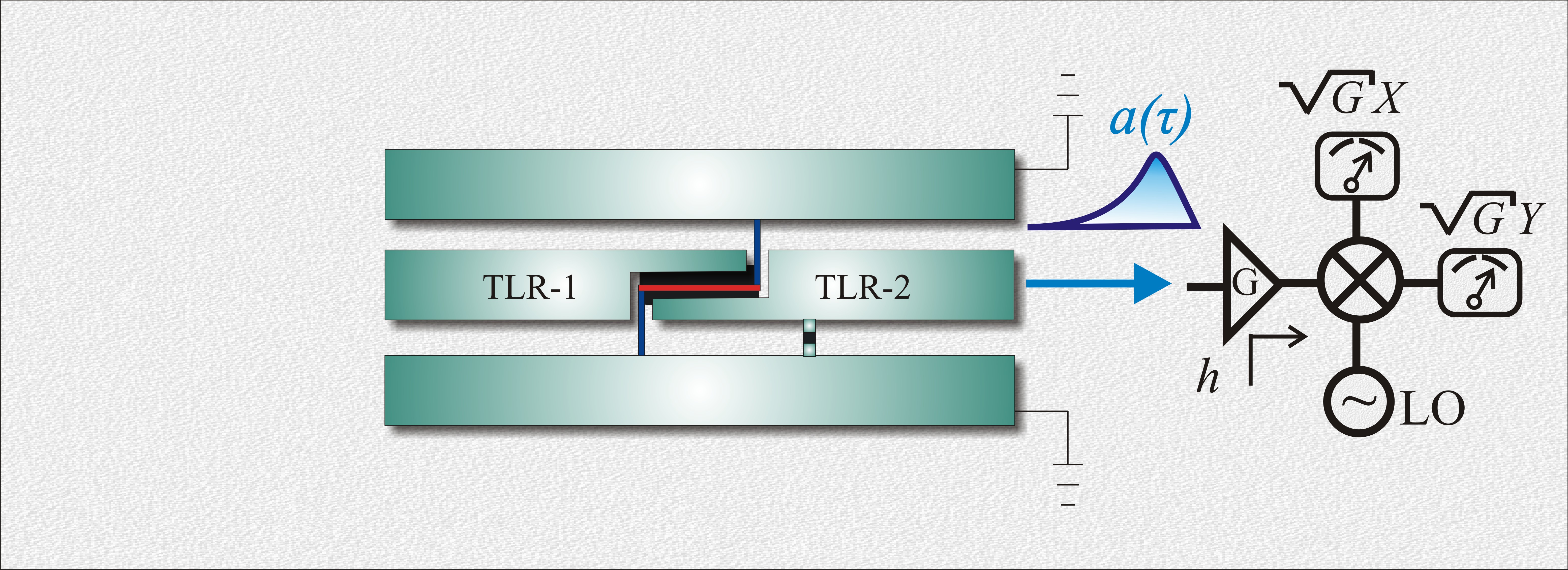}\\(d)\\
\caption{In the proposed experiment, the capacitive coupling between two TLRs is mediated by the vibration of a NEMS. The TLR-2, on the right is directly coupled with a Josephson Junction (JJ) for noise reduction. The pulsed measurement sequence is given as follows: (a) \textit{Step one.} Initial state preparation $\rho(0)$; (b) \textit{Step two.} Application of a pulse in TLR-2 by a coherent pulse  source of generating $\rho(t_{1})$; (c) \textit{Step three.} Application of a pulse in TLR-1 by another coherent pulse source generating $\rho(\tau);$ (d) \textit{Step four}. The output of the TLR-2 is directed to a microwave beam splitter (a hybrid). The two outputs are then amplified and run through separate IQ mixers. The four output currents from two IQ mixers can be correlated in various ways. The particlar correlation employed here gives access directly to all the normally ordered moments of the cavity field at the start of the measurement. The cross correlations are computed after all the field has been detected and the TLR-2 returned to the vacuum state ready for the next pulse. (Note: the pulse applied in step 3 does not interact with  the JJ, for not producing sufficiently large charging energy.)}
\label{method}
\end{figure*}

We consider the coupling of two Transmission Line Resonators (TLRs) mediated by a mechanical resonator as considered in Ref. \cite{0} and represented in Fig. \ref{method}. By assuming the regime of rapid mechanical oscillations at the GHz scale, and that the TLR-1, being in a undepleted regime, is treated classically, the interaction Hamiltonian between the NEMS and the field in the TLR-2 is given by 
\begin{equation}
H_{I}=-\hbar\alpha(t)b^{\dag}b(a+a^{\dag}),
\label{Hamilnonian1}
\end{equation}
where $\alpha(t)$ is proportional to the microwave amplitude in the TLR-1 \cite{0} and $a$($a^\dagger$) is the annihilation(creation) operator of the normal mode in the TLR-2. $b$($b^\dagger$) is the mechanical resonator annihilation(creation) operator. As one can see in this interaction Hamiltonian the phonon number operator $b^\dagger b$ is a nondemolition variable. In addition the field in this remaining TLR is affected by the presence of a Josephson Junction (See Fig. \ref{method}), whose effect (as demonstrated in the Appendix) is to induce quadrature squeezing \cite{1,yurke1989observation,2,3,4,5,fnori} through a  parametric term \begin{equation}
H_{p}= -i\hbar\frac{\gamma}{2}(a^{2}-a^{\dag2}).
\label{Hamiltonian2}
\end{equation}
Without loss of generality we take $\gamma=\beta\chi^{(2)}/2$ as being real and positive, where $\chi^{(2)}$ is the nonlinear susceptibility and $\beta$ is the amplitude the coherent pulse source in the TLR-2.  We shall consider a pulsed measurement   scenario \cite{6,Bozyigit2010} where the
interaction in $H_I$ is turned on and off rapidly. Note that the unconditional phonon number operator is a {quantum non-demolition variable. Moreover given that $[b^\dagger b, H_p+H_I]=0$, back-action evading is guaranteed and therefore $b^\dagger b$ can be repeteadly measured without the back-action noise \cite{Braginsky1980}. Therefore the phonon number statistics is not changed from pulse to pulse. The detailed process of this pulsed measurement is described in what follows. For a detailed description of the elements involved forlinear detection please see Refs. \cite{6,Bozyigit2010}.}

{The experiment is carried out according to Fig. \ref{method}. The Josephson Junction acts as a source of coherent pulses, with time interval $t_{1}$, on the TLR-2
In sequence, a coherent pulse  is applied to  TLR-1, with amplitude $\alpha_{1}$ and time interval $t_{2}$. The radiation field there is treated classically, as a result of the undepleted regime \cite{0}. After these pulses, the density operator is given by
\begin{eqnarray}
\rho(\tau=t_{1}+t_{2})&=&e^{-iH_{I}t_{2}}e^{-iH_{p}t_{1}}\rho(0)e^{iH_{p}t_{1}}e^{iH_{I}t_{2}}.
\end{eqnarray}
Initially, the TLR-2  is prepared in a vacuum state, while the NEMS, which oscillates due to thermal excitation, in equilibrium is in a thermal state with mean phonon number $N$, 
\begin{eqnarray}
\rho(0)=\sum^{\infty}_{n=0}P(n)\left|n\right\rangle\left\langle n\right|_{b}\otimes\left|0\right\rangle\left\langle 0\right|_{a},
\label{rho0}
\end{eqnarray}
where $P(n)=N^{n}/(N+1)^{n+1}$ is the thermal phonon number distribution in the NEMS, being $N=\left(\exp(\hbar\nu/k_{B}T)-1\right)^{-1}$ its thermal number. Given the very shorts pulses with time intervals $\tau=t_{1}+t_{2}$, the evolution can be approximated as above to give\begin{equation}
\rho(\tau)=\sum^{\infty}_{n=0}P(n)\left|n\right\rangle\left\langle n\right|_{b}\otimes\left|\alpha_n,\gamma\right\rangle\left\langle \alpha_n,\gamma\right|_{a},
\label{rhotau}
\end{equation}
with $\alpha_{n}(\tau)=i\,n\int^{t_{2}}_{0}\alpha(t)dt\equiv i\,nA$, where $A$ is the area of the interaction pulse, and $\left|\alpha_n,\gamma\right\rangle_{a}$ is a coherent squeezed state given by $\left|\alpha_n,\gamma\right\rangle_{a}=\mathcal{D}(\alpha_{n})\mathcal{S}(\gamma)\left|n,0\right\rangle$, where $\mathcal{D}(\alpha_{n})=\exp\left[inA(a+a^{\dag})\right]$ is the displacement operator conditioned on the mechanical resonator phonon excitation number and $\mathcal{S}(\gamma)=\exp\left[(\gamma t_{1}/2)(a^{\dag2}-a^{2})\right]$ is the squeezing operator. 
}

At the end of a pulse the reduced cavity field state is described by a mixture of squeezed coherent states, which are distributed according to the mechanical resonator thermal phonon number distribution $P(n)$.
The field in TLR-2 is dumped  out through the dual IQ mixer measurement scheme of Ref. \cite{6,Bozyigit2010}, detailed in Fig. (\ref{method}). In
this scheme, the TLR-2 output is directed to a microwave beam splitter (a hybrid coupler). The two outputs are
then amplified and run through separate IQ mixers. The four output currents from the IQ mixers can be correlated
in various ways after proceeding to linear detectors \cite{6}. One particular cross correlation gives access directly to all the normally ordered moments of the
TLR-2 field at the initial time of the measurement.
At the end of each pulse a measurement is performed on the output of the second TLR through complexes envelopes $S_{c}=-a(\tau)$ and $S_{d}=ia(\tau)$. 
{It is important to remark that the criterion for back-action-evasion for $b^\dagger b$, mainly $[b^\dagger b, H_p+H_I]=0$, is obeyed in every detection step.}

These pulsed measurements allow that the 
cross-correlations be computed after all the field has been detected and the TLR-2 returned to the vacuum state ready for the next pulse.
For each individual pulsed measurement the NEMS is to be found on an specific phonon number excitation $m$ according to the thermal distribution $P(m)$. The post-selected state of the TLR-2  field conditioned on this instantaneous (at the time of measurement $\tau$) NEMS excitation  $m$ is given by
\begin{eqnarray}
\rho^{(m)}_a(\tau)&=&\frac{\Tr_{b}\left\{\Pi_{m}\rho(\tau)\Pi_{m}\right\}}{\Tr_{ab}\left\{\Pi_{m}\rho(\tau)\Pi_{m}\right\}},
\end{eqnarray}
where $\Pi_{m}=\left|m\right\rangle\left\langle m\right|_{b}$, $\rho(\tau)$ is given by Eq. (\ref{rhotau}), and ${\Tr_{ab}\left\{\Pi_{m}\rho(\tau)\Pi_{m}\right\}}=P(m)$ is the probability to obtain $m$. 
Therefore the 
 expected value of the phase quadrature in each measurement  
 \be\left\langle Y\right\rangle_{(m)}(\tau)=\left\langle \alpha_{m},\gamma\right|Y\left|\alpha_{m},\gamma\right\rangle,\ee
  has a variance
   \be\left\langle (\Delta Y)^{2}\right\rangle_{(m)}(\tau)=e^{-2\gamma t_{1}}.\label{varm}\ee
 The TLR-2 field pre-selected state $\rho_{a}(\tau)=\sum_{m}P(m)\rho_{a}^{(m)}(\tau)$ average over all the quadrature measurements as $\left\langle Y\right\rangle(\tau)=\sum_{m}P(m)\left\langle Y\right\rangle_{(m)}(\tau)$ providing the phonon number (and consequently the temperature) measurement. 
 
\begin{figure}[!h]
\includegraphics[scale=0.5]{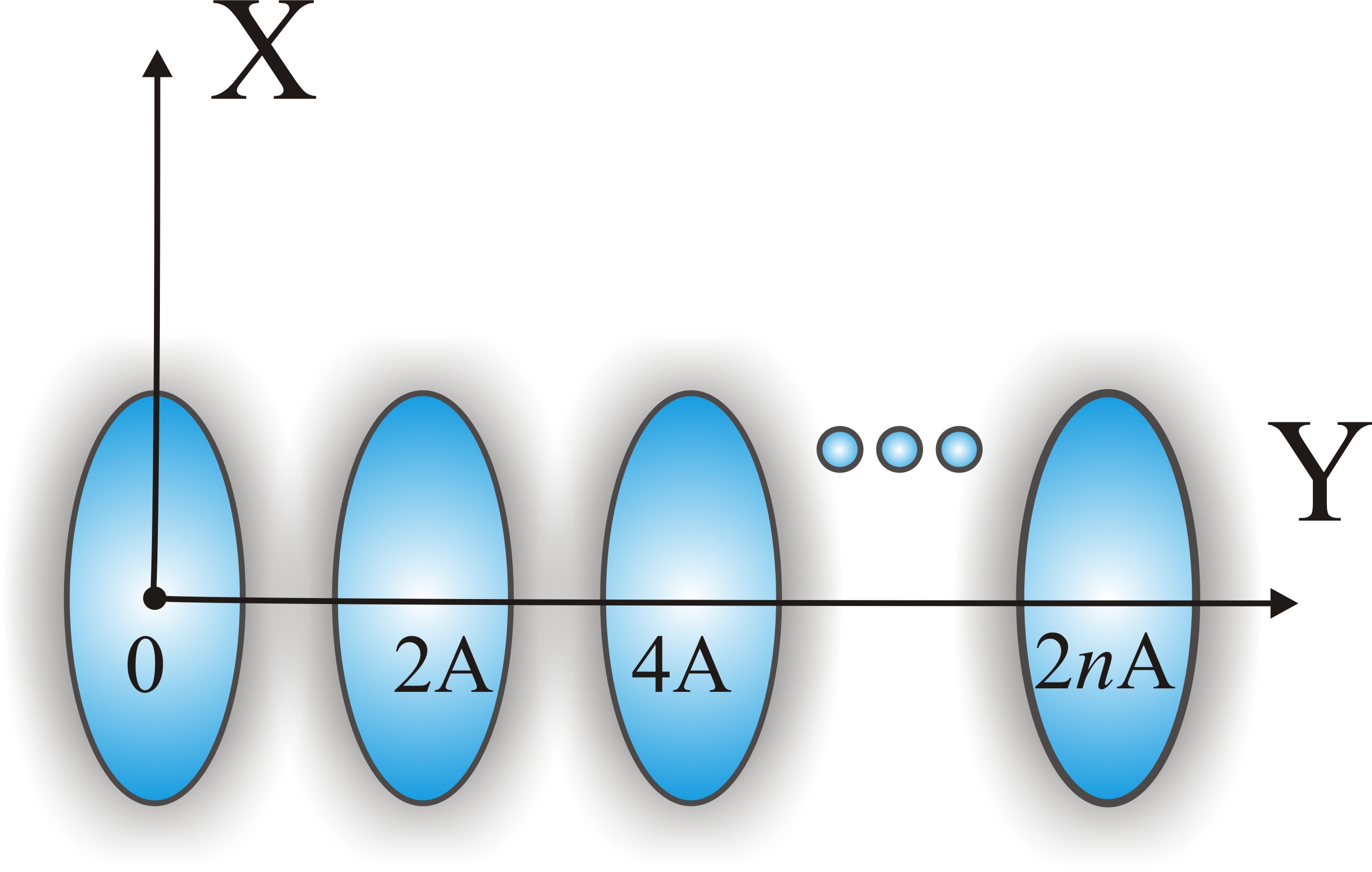}
\caption{Phase space  of the TLR-2 in $t=\tau$ for $\gamma t_{1}>-ln[\sqrt{2A}]$. The minor (major) axis of the ellipses represent $\left\langle (\Delta Y)^{2}\right\rangle^{(m)}(\tau)=e^{-2\gamma t_{1}}$ ($\left\langle (\Delta X)^{2}\right\rangle^{(m)}(\tau)=e^{2\gamma t_{1}}$), the variance of each measurement of the quadrature  $Y$ ($X$). The separation of the ellipses along $Y$ is equal to pulse the interaction area $2A$.}
\label{spacephase}
\end{figure}

 The computation of the total variance on $Y$ depends on the variance of the thermal NEMS by adding the variance of each individual measurement, and is dependent on the distinguishability of each individual detection. The squeezing enables the disjunction of the individual conditioned TLR-2 field detections, allowing to attribute the correct excitation $m$ influencing $\rho_a^{(m)}(\tau)$, as depicted in the phase space in Fig. \ref{spacephase}. Therefore the squeezing improves the statistical resolution, being necessary that $\gamma t_{1}>-ln[\sqrt{2A}]$ for that to be ensured.  The ellipses are centered at $n 2A$, $n=0,1,2,...$ , and their minor and major axes represent the fluctuation of the measurement of each quadrature.  Without the squeezing an incorrect attribution could be given to $m$ through the fluctuation $\left\langle (\Delta Y)^{2}\right\rangle_{(m)}(\tau)=e^{-2\gamma t_{1}}$.

 The resulting average over many pulsed measurements is given by the quadratures
\begin{eqnarray}
\left\langle Y\right\rangle(\tau)=\Tr\left\{i\left(a^{\dag}-a\right)\rho(\tau)\right\}=2AN,
\label{temperature}
\end{eqnarray}
and
\begin{eqnarray}
\left\langle X\right\rangle(\tau)=\Tr\left\{\left(a+a^{\dag}\right)\rho(\tau)\right\}=0.
\label{notemperature}
\end{eqnarray}
Therefore, by averaging over many identical pulses we can reconstruct the mean phonon number of the
NEMS, $N$, and thus deduce its temperature, $T=\hbar \nu/[k_B \ln{(N^{-1}+1)}]$ . In a similar fashion we could also measure two-time correlation functions for the mechanical resonator. Instead here we focus on the quadratures variance and the increasing of accuracy in determining the phonon statistics of the mechanical resonator due to the presence of the squeezing term in Eq. (\ref{Hamiltonian2}).

Through the pulsed measurement the variance of the quadrature $Y$ is given by
\begin{equation}
\left\langle (\Delta Y)^{2}\right\rangle=4A^{2}N\left(N+1\right)+
 \left\langle (\Delta Y)^{2}\right\rangle_{(m)}(\tau),
\label{varianceY}
\end{equation}
being $\left\langle (\Delta Y)^{2}\right\rangle_{(m)}(\tau)$, the variance of each individual measurement given by Eq. (\ref{varm}).
$N(N+1)$  appearing in Eq. (\ref{varianceY}) is the variance of the thermal distribution of the NEMS.
The squeezing induced by the Josephson Junction allows a reduction of the noise on each measurement, in contrast with the situation without the Josephson Junction, for which $\left\langle (\Delta Y)^{2}\right\rangle_{(m)}(\tau)=1$. In the limit of large squeezing the relative uncertainty $\widetilde{\Delta Y}\equiv\sqrt{\left\langle (\Delta Y)^{2}\right\rangle}/\left\langle Y\right\rangle$ is approximately given by
\begin{equation}
\widetilde{\Delta Y}\approx \sqrt{1+N^{-1}}, \end{equation}
being independent on the pulse area and reapidly reaching the threshold  $1$ with increasing $N$.



\begin{figure}[!h]
\includegraphics[scale=0.20]{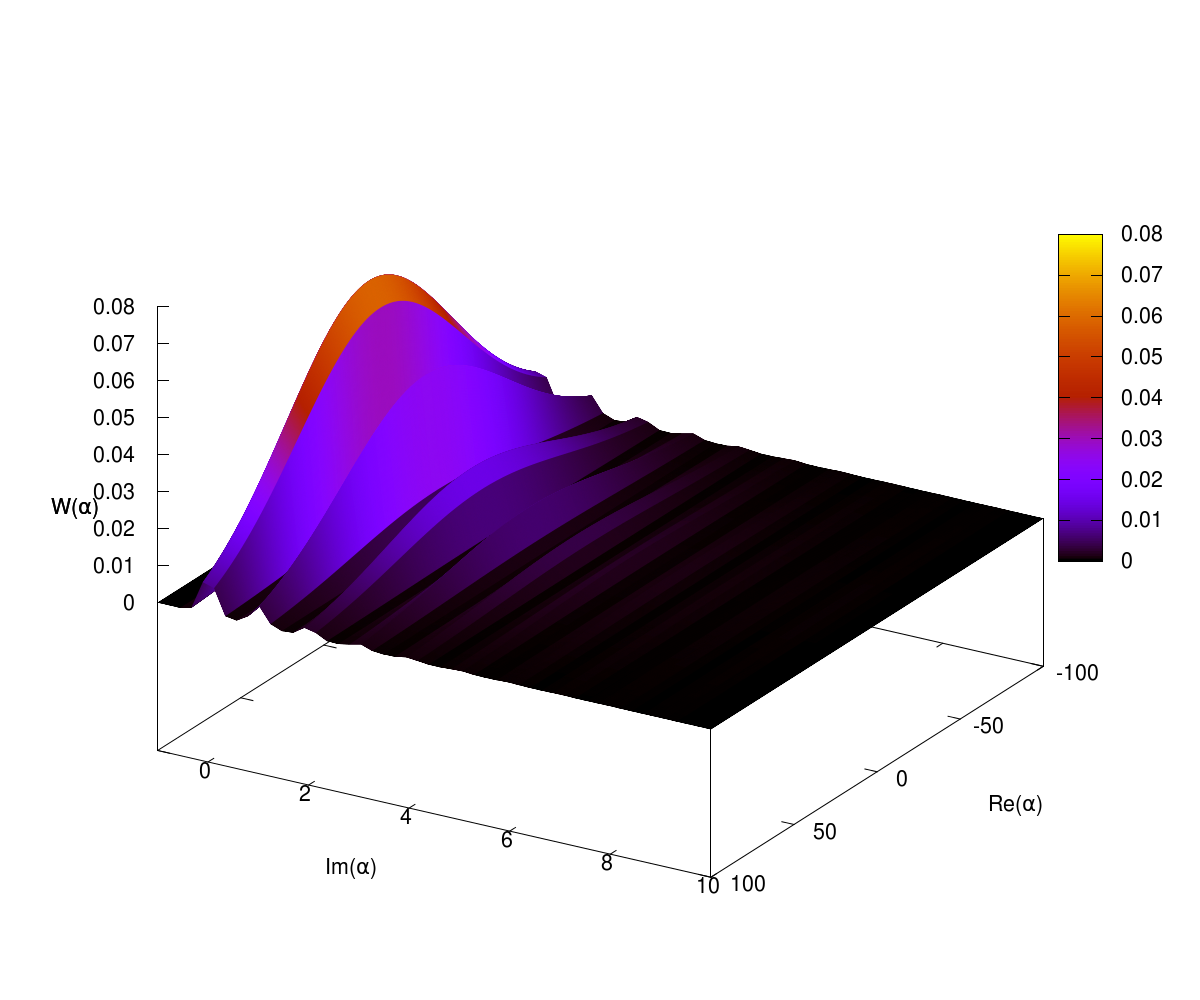}\\(a)\\
\includegraphics[scale=0.30]{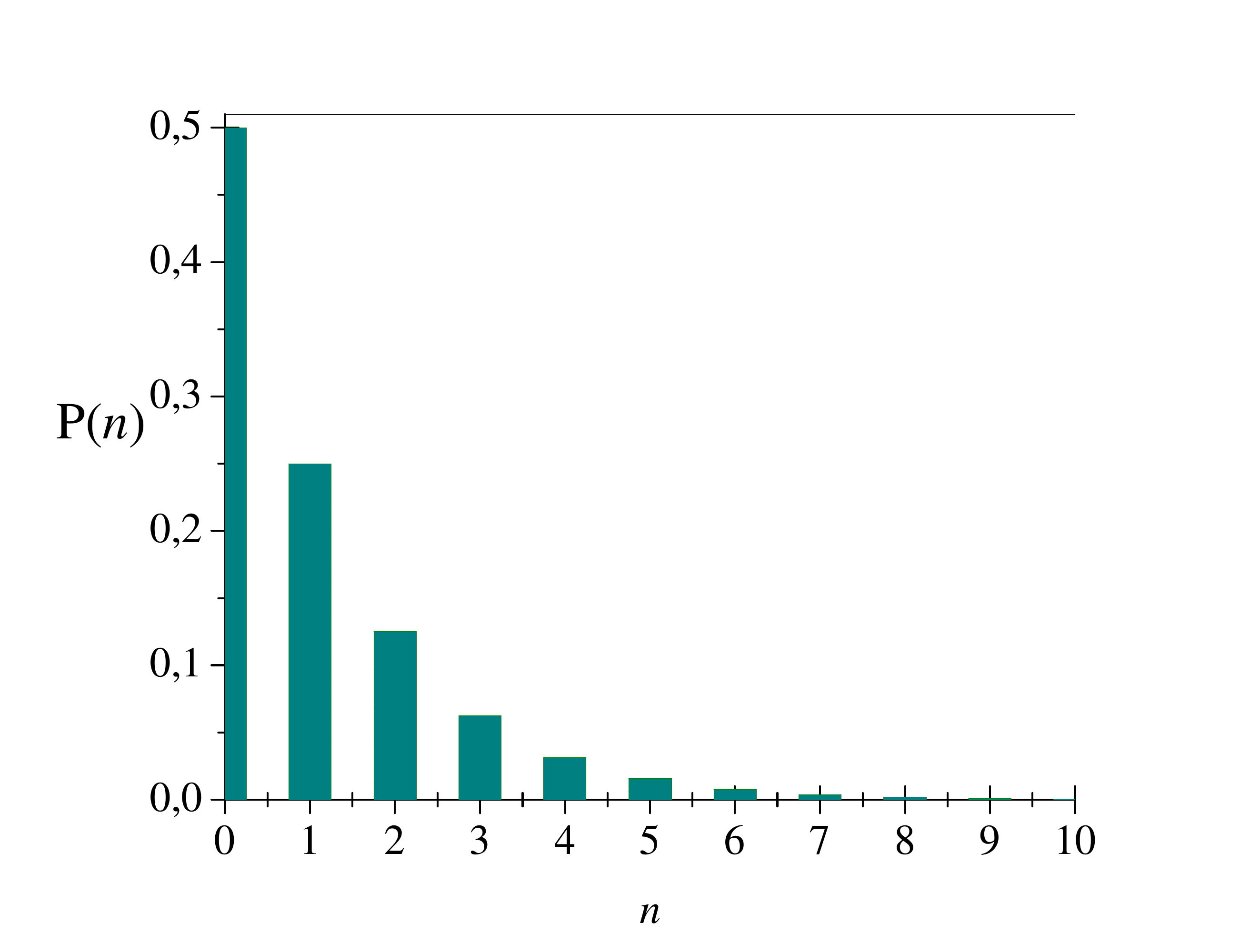}\\(b)\\
\caption{(a) Wignerdistibution for the TLR-2 field, as given in Eq. (\ref{w3dh}) for  $N=1$, $A=1$ and $e^{2\gamma t_{1}}=50$. Integrating the Wigner distribution with respect to variable $Re(\alpha)$, $\mathcal{P}=\int dRe(\alpha)\mathcal{W}(\alpha)$, we obtain the visualization of the NEMS thermal phonon distribuiton $P(n)$ depicted \cite{nation2011qutip} in the histogram (b).}
\label{Wignerandhistogram}
\end{figure}

Extending the previous procedure for measurement of the TLR-2 field ordered moments we can reconstruct the Wigner quasiprobability distribution \cite{,eichler2011experimental,haroche2013exploring}
\begin{equation}
\mathcal{W}(\alpha)=\sum_{m,n}\int d^2\lambda \frac{\langle (a^\dagger)^n a^m\rangle(-\lambda^*)^m\lambda^n}{\pi^2 n!m!} e^{-\frac{1}{2}|\lambda|^2+\alpha\lambda^*-\alpha^* \lambda},
\end{equation}
for the field state, giving
\begin{equation}
\mathcal{W}(\alpha)=\sum_{n}\frac{P(n)}{2\pi}e^{-\frac{1}{2}Re^2(\alpha)e^{-2\gamma t_{1}}-\frac{1}{2}\left[Im(\alpha)-\alpha_{n}\right]^{2}e^{2\gamma t_{1}}}.
\label{w3dh}
\end{equation}
The signature of the NEMS phonon distribution $P(n)$ is clearly imprinted in the Wigner distribution profile  in Fig.(\ref{Wignerandhistogram}) and can accessed through the marginal distibution $\mathcal{P}=\int dRe(\alpha)\mathcal{W}(\alpha)$.

\section{Discussion}
Detection of thermodynamical quantities associated to nanoscopic movement is important in many aspects. Particularly, the detection of the corresponding temperature of such tiny devices is relevant for physical characterization \cite{Chang2007,Stace2010} and latter usage in further applications. Here we have proposed a mechanism to measure the average number of phonons of a NEMS through a pulsed non-demolition detection scheme. Since here the NEMS phonons are due to thermal excitation, a direct way to characterize the temperature of the device as well as it statistical properties in a non-demolition way is derived.
The proposed scheme to measure the temperature of the NEMS is experimentally feasible with nowadays technology. The interaction between semiconducting NEMS \cite{World,Preport,Ptoday,roukes2001plenty,Lahaye,ekinci2005electromechanical} and superconducting TLR \cite{fabric,schoelkopf2008wiring,devoret2007circuit} has already been carried out \cite{regal2008measuring,zhou2013slowing,Tlrnems,woolley2008nanomechanical,teufel2009nanomechanical}. Experiments with The IQ mixer measuring method were also performed \cite{Bozyigit2010}. Therefore we expect that the presented scheme be readily implemented. {The mechanical resonator is assumed to be at thermal equilibrium at the measurement stage, and therefore dissipative effects, which are strong for those devices play a key role for reaching this equilibrium. That is to be reached prior the sequence of pulses of the detection scheme, otherwise it would be expected a continuous decrease of the phonon number, which would be typical in a transient regime. The more relevant dissipative effect comes in fact from the superconducting charge qubit decayng time\cite{DeSaNeto2011} (See \cite{asp2014} for a compreensive review of some recent experimental numbers). At the frequencies of GHz necessary for the present porposal the charge qubit decay will affect the squeezing of the TLR 2 field, and therefore will affect the accuracy of the detection process. However given the ability to perform short and fast pulses in those devices \cite{6} we do not expect it to be detrimental for the present proposal. A more detailed investigation on that is required.} 

\section*{acknowledgements}

OPSN work is supported in part by CAPES. MCO acknowledges support by FAPESP and CNPq through the National Institute for Science and Technology on Quantum Information and the Research Center in
Optics and Photonics (CePOF). GJM acknowledges the support of the Australian Research Council CE110001013. OPSN is grateful to L. D. Machado, S. S. Coutinho, K. M. S. Garcez, J. Lozada-Vera, A. Carrillo and F. Nicacio for helpful discussions.

\appendix*

\section{Derivation of Hamiltonian (\ref{Hamiltonian2}).} \label{secmet}

\begin{figure}[!h]
\includegraphics[scale=0.5]{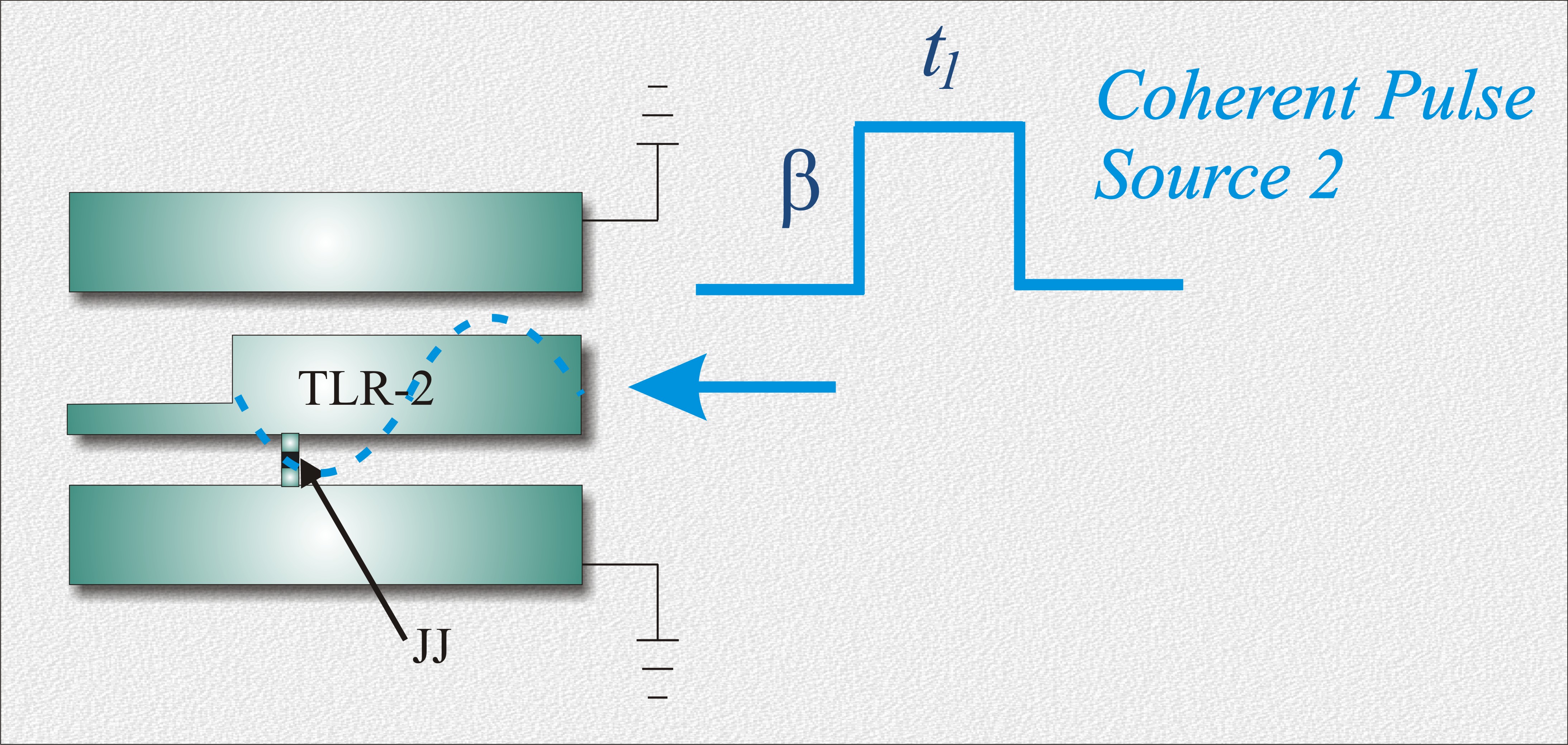}\\(a)\\
\includegraphics[scale=0.5]{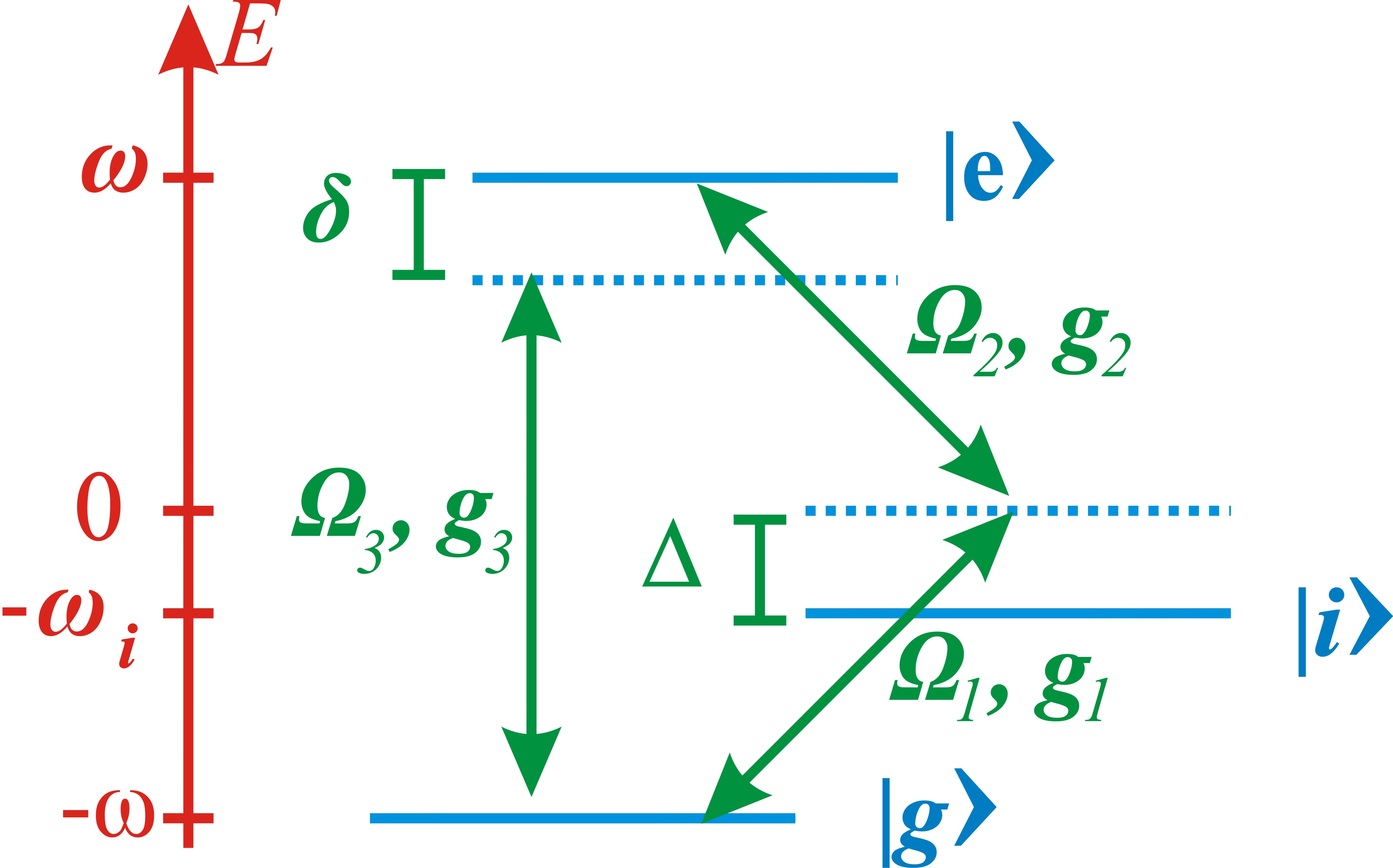}\\(b)\\
\caption{(a) Scheme for generation of a parametric oscillator, via the pulse applied to the TLR-2, coupled to an artificial three-level atom; (b) Energy-level diagram, observing the Hamiltonian (\ref{Hamiltonianzero}), (\ref{Hamiltonianofinteraction}) and  $\Delta=-(\Omega_{i}-\Omega_{g}-\Omega_{1})=\Omega_{e}-\Omega_{i}-\Omega_{2}$ (with $\Omega_{g}=-\omega$, $\Omega_{e}=\omega$ and $\Omega_{i}=-\omega_{i}$) and $\Omega_{3}=\Omega_{e}-\Omega_{g}-\delta$.}
\label{POthree}
\end{figure}

We consider the model in figure (\ref{POthree}), where the three lowest Josephson Junction charge states form a three-level artificial atom is in a configuration where the ground ($\left|g\right\rangle$) and excited ($\left|e\right\rangle$) states are separated by an intermediary atomic level ($\left|i\right\rangle$). The modes with frequency $\Omega_{1}$, $\Omega_{2}$ and $\Omega_{3}$ are coupled to the transitions $\left|g\right\rangle \leftrightarrow \left|i\right\rangle$, $\left|i\right\rangle \leftrightarrow \left|e\right\rangle$ and $\left|g\right\rangle \leftrightarrow \left|e\right\rangle$ with coupling constants $g_{1}$, $g_{2}$ and $g_{3}$, respectively.

The Hamiltonian in the rotating wave approximation is 
\begin{eqnarray}
H=H_{0}+V,
\label{HPrwa}
\end{eqnarray}
being
\begin{equation}
H_{0}=\hbar\sum_{j=1}^3\Omega_{j}a_{j}^{\dag}a_{j}+\hbar\Omega_{e}\sigma_{ee}+\hbar\Omega_{i}\sigma_{ii}+\hbar\Omega_{g}\sigma_{gg},
\label{Hamiltonianzero}
\end{equation}
and
\begin{equation}
V=\hbar\left(g_{1}a_{1}^{\dag}\sigma_{ig}+g_{2}a_{2}^{\dag}\sigma_{ei}+g_{3}a_{3}^{\dag}\sigma_{eg}+h.c.\right),
\label{Hamiltonianofinteraction}
\end{equation}
where $\sigma_{jk}=\left|j\right\rangle\left\langle k\right|$ is the atomic transition for $j,k=g,i,e$. Writing $H$ in the interaction picture, by applying  $U_{0}(t)=e^{-iH_{0}t/\hbar}$, and transforming it through $U_{1}=e^{-i\Delta t(\sigma_{ee}+\sigma_{gg})}$, we obtain
\begin{eqnarray}
\frac{\tilde{H}}{\hbar}&=&-\Delta\left(\sigma_{ee}+\sigma_{gg}\right)\nonumber\\
&&+g_{1}a_{1}\sigma_{gi}+g_{2}a_{2}\sigma_{ie}+g_{3}a_{3}\sigma_{ge}\nonumber\\
&&+g_{1}^{*}a_{1}^{\dag}\sigma_{ig}+g_{2}^{*}a_{2}^{\dag}\sigma_{ei}+g_{3}^{*}a_{3}^{\dag}\sigma_{eg}.
\label{Hamiltoniantilde}
\end{eqnarray}
Therefore, the Heisenberg equations of motion for the coherences $\sigma_{gi}$ and $\sigma_{ie}$ are
\begin{equation}
i\dot{\sigma}_{ig}=-\Delta\sigma_{ig}+g_{1}a_{1}\left(\sigma_{ii}-\sigma_{gg}\right)+g_{3}a_{3}\sigma_{ie}-g_{2}^{*}a_{2}^{\dag}\sigma_{eg},
\end{equation}
and
\begin{equation}
i\dot{\sigma}_{ie}=-\Delta\sigma_{ie}+g_{2}^{*}a_{2}^{\dag}\left(\sigma_{ii}-\sigma_{ee}\right)+g_{3}^{*}a_{3}^{\dag}\sigma_{ig}-g_{1}a_{1}\sigma_{ge}.
\end{equation}
Assuming that initially the artificail atom is prepared in the state $|i\rangle$ and that both $\Delta$ and $\delta$ and large enough so that the states $|e\rangle$ and $|g\rangle$ are not significantly populated so that $\sigma_{ii}=1$, in the limit where $\Delta\gg\left|g_{1}\right|,\left|g_{2}\right|,\left|g_{3}\right|,\delta$, we have the adiabatic solution with $\dot{\sigma}_{ig}=\dot{\sigma}_{ie}=0$. Next, the expression of $\sigma_{gi}$ and $\sigma_{ie}$, we can consider $\Delta^{2}-g_{3}^{2}a_{3}^{\dag}a_{3}\cong\Delta^{2}$, the terms with $\sigma_{gg}$, $\sigma_{ee}$, $\sigma_{ge}$ and $\sigma_{eg}$ can be neglected, resulting in the solution
\begin{eqnarray}
\sigma_{ig}&\cong&\frac{g_{1}}{\Delta}a_{1}+\frac{g_{2}^{*}g_{3}}{\Delta^{2}}a_{2}^{\dag}a_{3}
\label{solutionig}
\end{eqnarray}
and
\begin{eqnarray}
\sigma_{ie}&\cong&\frac{g_{2}^{*}}{\Delta}a_{2}^{\dag}+\frac{g_{1}g_{3}^{*}}{\Delta^{2}}a_{1}a_{3}^{\dag}.
\label{solutionie}
\end{eqnarray}
In this way, preparing the artificial atom in intermediate state $\left|i\right\rangle$, and replacing (\ref{solutionig}) and (\ref{solutionie}) expressions in Hamiltonian (\ref{Hamiltoniantilde}), we obtain the effective Hamiltonian of the form
\begin{equation}
\frac{\tilde{H}_{eff}}{2\hbar}=\frac{g_{1}^{2}}{\Delta}a_{1}^{\dag}a_{1}+\frac{g_{2}^{2}}{\Delta}a_{2}^{\dag}a_{2}+\frac{g_{1}g_{2}g_{3}^{*}}{\Delta^{2}}a_{1}a_{2}a_{3}^{\dag}+\frac{g_{1}g_{2}g_{3}}{\Delta^{2}}a_{1}^{\dag}a_{2}^{\dag}a_{3},
\end{equation}
being $g_{1}$ and $g_{2}$ real. Now, applying the transformation $U_{2}=e^{-it(g_{1}^{2}a_{1}^{\dag}a_{1}+g_{2}^{2}a_{2}^{\dag}a_{2})/\Delta}$, $a_{3}=i\beta$ being a coherent pulse source of amplitude $\beta$, and $g_{3}\rightarrow\mathcal{G}_{3}e^{-i\delta t}$ so that its frequency is adjusted to $\delta=(g_{1}^{2}+g_{2}^{2})/\Delta$, we obtain
\begin{equation}
H_{p}=-i\hbar\frac{\gamma}{2}(a^{2}-a^{\dag2}),
\end{equation}
with $a_{1}=a_{2}=a$, $\Omega_{1}=\Omega_{2}=\omega$, $\gamma=4g_{1}g_{2}\mathcal{G}_{3}\beta/\Delta^{2}$. Note that in this case, the effective frequency is satisfied for $\Omega_{3}=2\omega-(g_{1}^{2}+g_{2}^{2})/\Delta$.


\bibliographystyle{apsrev4-1}
\bibliography{ref}

\end{document}